\documentstyle[twoside,fleqn,espcrc2]{article}

\newcommand{\AmS}{{\protect\the\textfont2
  A\kern-.1667em\lower.5ex\hbox{M}\kern-.125emS}}

\begin{document}
\begin{titlepage}
\hskip 12cm
\vbox{\hbox{hep-th/9709113}\hbox{CERN-TH/97-226}\hbox{September, 1997}}
\vfill
\begin{center}
{\LARGE {Duality, Central Charges and Entropy of\\
 Extremal BPS Black Holes }}\\
\vskip 1.5cm
{  {\bf Laura Andrianopoli$^1$,
Riccardo D'Auria$^2$ and
Sergio Ferrara$^3$ }} \\
\vskip 0.5cm
{\small
$^1$ Dipartimento di Fisica, Universit\'a di Genova, via Dodecaneso 33,
I-16146 Genova\\
and Istituto Nazionale di Fisica Nucleare (INFN) - Sezione di Torino, Italy\\
\vspace{6pt}
$^2$ Dipartimento di Fisica, Politecnico di Torino,\\
 Corso Duca degli Abruzzi 24, I-10129 Torino\\
and Istituto Nazionale di Fisica Nucleare (INFN) - Sezione di Torino, Italy\\
\vspace{6pt}
$^3$ CERN Theoretical Division, CH 1211 Geneva 23, Switzerland
}
\end{center}
\vfill
\vskip 2cm
\begin{center}
 Talk given by
  S. Ferrara at the \\
``{\it STRINGS '97}'' Conference,\\  16-21 June 1997,
 Amsterdam, The Netherlands
\end{center}
\vskip 2cm
\begin{center} {\bf Abstract}
\end{center}
{
\small
We report on some general results on the physics of extremal BPS black holes
in four and five dimensions.
The duality-invariant entropy-formula for all $N>2$ extended supergravities
is derived.
Its relation with the fixed-scalar condition for the black-hole ``potential energy''
 wich extremizes the BPS mass is obtained.
BPS black holes preserving different fractions of supersymmetry are classified
 in a U-duality invariant set up.
The latter deals with different orbits of the fundamental representations of the
exceptional groups $E_{7(7)}$ and $E_{6(6)}$.
We comment upon the interpretation of these results
in a string and M-theory framework.
}
\vspace{2mm} \vfill \hrule width 3.cm
{\footnotesize
\noindent
$^*$ Work supported in part by EEC under TMR contract ERBFMRX-CT96-0045
 (LNF Frascati,
Politecnico di Torino and Univ. Genova) and by DOE grant
DE-FGO3-91ER40662}
\end{titlepage}

\hyphenation{author another created financial paper re-commend-ed}

\title{Duality, Central Charges and Entropy of Extremal BPS Black Holes}

\author{L. Andrianopoli\address{Dipartimento di Fisica, Universit\`a di Genova, via Dodecaneso 33,
I-16146 Genova\\
and Istituto Nazionale di Fisica Nucleare (INFN) - Sezione di Torino, Italy },
 R. D'Auria\address{Dipartimento di Fisica, Politecnico di Torino,\\
 Corso Duca degli Abruzzi 24, I-10129 Torino\\
and Istituto Nazionale di Fisica Nucleare (INFN) - Sezione di Torino, Italy} %
and S. Ferrara\address{CERN Theoretical Division, CH-1211 Geneva 23, Switzerland }}

\newcommand{\ft}[2]{{\textstyle\frac{#1}{#2}}}
\newcommand{\QED}{{\hspace*{\fill}\rule{2mm}{2mm}\linebreak}}
\def\dop{{\rm d}\hskip -1pt}
\def\bfone{\relax{\rm 1\kern-.35em 1}}
\def\bfzero{\relax{\rm I\kern-.18em 0}}
\def\inbar{\vrule height1.5ex width.4pt depth0pt}
\def\IC{\relax\,\hbox{$\inbar\kern-.3em{\rm C}$}}
\def\ID{\relax{\rm I\kern-.18em D}}
\def\IF{\relax{\rm I\kern-.18em F}}
\def\IK{\relax{\rm I\kern-.18em K}}
\def\IH{\relax{\rm I\kern-.18em H}}
\def\II{\relax{\rm I\kern-.17em I}}
\def\IN{\relax{\rm I\kern-.18em N}}
\def\IP{\relax{\rm I\kern-.18em P}}
\def\IQ{\relax\,\hbox{$\inbar\kern-.3em{\rm Q}$}}
\def\IR{\relax{\rm I\kern-.18em R}}
\def\IG{\relax\,\hbox{$\inbar\kern-.3em{\rm G}$}}
\font\cmss=cmss10 \font\cmsss=cmss10 at 7pt
\def\ZZ{\relax\ifmmode\mathchoice
{\hbox{\cmss Z\kern-.4em Z}}{\hbox{\cmss Z\kern-.4em Z}}
{\lower.9pt\hbox{\cmsss Z\kern-.4em Z}}
{\lower1.2pt\hbox{\cmsss Z\kern-.4em Z}}\else{\cmss Z\kern-.4em
Z}\fi}
\def\a{\alpha} \def\b{\beta} \def\d{\delta}
\def\e{\epsilon} \def\c{\gamma}
\def\G{\Gamma} \def\l{\lambda}
\def\L{\Lambda} \def\s{\sigma}
\def\cA{{\cal A}} \def\cB{{\cal B}}
\def\cC{{\cal C}} \def\cD{{\cal D}}
\def\cF{{\cal F}} \def\cG{{\cal G}}
\def\cH{{\cal H}} \def\cI{{\cal I}}
\def\cJ{{\cal J}} \def\cK{{\cal K}}
\def\cL{{\cal L}} \def\cM{{\cal M}}
\def\cN{{\cal N}} \def\cO{{\cal O}}
\def\cP{{\cal P}} \def\cQ{{\cal Q}}
\def\cR{{\cal R}} \def\cV{{\cal V}}\def\cW{{\cal W}}
%
%
\def\crr{\crcr\noalign{\vskip {8.3333pt}}}
\def\tilde{\widetilde}
\def\bar{\overline}
\def\us#1{\underline{#1}}
\let\shat=\hat
\def\hat{\widehat}
\def\hyp{\vrule height 2.3pt width 2.5pt depth -1.5pt}
\def\square{\mbox{.08}{.08}}
\def\Coeff#1#2{{#1\over #2}}
\def\Coe#1.#2.{{#1\over #2}}
\def\coeff#1#2{\relax{\textstyle {#1 \over #2}}\displaystyle}
\def\coe#1.#2.{\relax{\textstyle {#1 \over #2}}\displaystyle}
\def\half{{1 \over 2}}
\def\shalf{\relax{\textstyle {1 \over 2}}\displaystyle}
\def\dag#1{#1\!\!\!/\,\,\,}
\def\to{\rightarrow}
\def\notin{\hbox{{$\in$}\kern-.51em\hbox{/}}}
\def\shdot{\!\cdot\!}
\def\ket#1{\,\big|\,#1\,\big>\,}
\def\bra#1{\,\big<\,#1\,\big|\,}
\def\equaltop#1{\mathrel{\mathop=^{#1}}}
\def\Trbel#1{\mathop{{\rm Tr}}_{#1}}
\def\inserteq#1{\noalign{\vskip-.2truecm\hbox{#1\hfil}
\vskip-.2cm}}
\def\attac#1{\Bigl\vert
{\phantom{X}\atop{{\rm\scriptstyle #1}}\phantom{X}}}
\def\exx#1{e^{{\displaystyle #1}}}
\def\del{\partial}
\def\delbar{\bar\partial}
\def\nex#1{$N\!=\!#1$}
\def\dex#1{$d\!=\!#1$}
\def\cex#1{$c\!=\!#1$}
\def\eg{{\it e.g.}} \def\ie{{\it i.e.}}

%
\def\cS{{\cal K}}
\def\IE{\relax{{\rm I\kern-.18em E}}}
\def\IGam{\relax{{\rm I}\kern-.18em \Gamma}}
\def\IGa{\IA}
\def\IA{\relax{\hbox{{\rm A}\kern-.82em {\rm A}}}}
\let\picfuc=\fp
\def\hata{{\shat\a}}
\def\hatb{{\shat\b}}
\def\hatA{{\shat A}}
\def\hatB{{\shat B}}
\def\bv{{\bf V}}
%


\def\twomat#1#2#3#4{\left(\begin{array}{cc}
 {#1}&{#2}\\ {#3}&{#4}\\
\end{array}
\right)}
\def\twovec#1#2{\left(\begin{array}{c}
{#1}\\ {#2}\\
\end{array}
\right)}

\begin{abstract}
We report on some general results on the physics of extremal BPS black holes
in four and five dimensions.
The duality-invariant entropy-formula for all $N>2$ extended supergravities
is derived.
Its relation with the fixed-scalar condition for the black-hole ``potential energy''
 wich extremizes the BPS mass is obtained.
BPS black holes preserving different fractions of supersymmetry are classified
 in a U-duality invariant set up.
The latter deals with different orbits of the fundamental representations of the
exceptional groups $E_{7(7)}$ and $E_{6(6)}$.
We comment upon the interpretation of these results
in a string and M-theory framework.
\end{abstract}

\maketitle


\section{Introduction}
In recent time, remarkable results have been obtained
 in the study of general properties
of BPS states both in supersymmetric gauge theories as well as in
 supersymmetric theories of gravity.
The latter are described by
 string theory and M-theory \cite{string} whose symmetry properties are encoded
in extended supergravity effective field theories.

Of particular interest are extremal black holes in four and five dimensions which
correspond to BPS saturated states \cite{black} and whose ADM mass depends,
 beyond the
quantized values of electric and magnetic charges, on the asymptotic value
 of scalars at infinity.
The latter describe the moduli space of the theory
Another physical relevant quantity, which depends only on quantized electric
 and magnetic charges,
 is the black hole entropy,
which can be defined macroscopically, through the Bekenstein-Hawking
 area-entropy relation
or microscopically, through D-branes techniques \cite{dbr} by counting
of microstates \cite{micros}.
It has been further realized that the scalar fields, independently of
their values
at infinity, flow towards the black hole horizon to a fixed value of pure
 topological
nature given by a certain ratio of electric and magnetic charges \cite{fks}.
These ``fixed scalars'' correspond to the extrema of the ADM mass
in moduli space while the black-hole entropy  is the  value of the
 squared
ADM mass at this point in $D=4$  \cite{feka1} \cite{feka2} and the power $3/2$
of the ADM mass in $D=5$.
In four dimensional theories with $N>2$, extremal black-holes preserving one supersymmetry
have the further property that all central charge eigenvalues other than the one
equal to the BPS mass flow to zero for ``fixed scalars''.
This is not true in $D=5$ because the charges transform in the
antisymmetric representation of $Usp(N)$ instead of $U(N)$ as in the
four dimensional cases.

The entropy formula turns out to be in all cases a U-duality invariant
expression
(homogeneous of degree two in $D=4$ and of degree $3/2$ in $D=5$)
built out of electric and magnetic charges and as such
can be in fact also computed through certain (moduli-independent) topological
quantities which only depend on the nature of the U-duality groups and the
appropriate representations
of electric and magnetic charges.
For example, in the $N=8$, $D=4$ theory  the entropy was shown to correspond  to the
unique quartic $E_7$ invariant built with its 56 dimensional representation
\cite{kall}.
In this report  we
intend to summarize further progress in this subject by  deriving, for all $N>2$
theories in $D=4,5$,
topological (moduli-independent) U-invariants constructed in terms of
(moduli-dependent)
 central charges and matter charges, and show that, as expected, they coincide
with $M^2_{ADM}$ or $M^{3/2}_{ADM}$ (in the case of the four and five dimensional
theories respectively),
computed at ``fixed scalars''.

The situation of black-hole backgrounds preserving more than $1/8$ of
the original  supersymmetry (32 charges)
is further explored.

Sections 2 and 3 deal with the four and five dimensional cases respectively,
section 4 describes the absolute duality invariants, section 5 describes BPS
states preserving more than one supersymmetry and section 6 gives a further classification
of BPS states in terms of duality orbits.

\section{Central charges, U-invariants and entropy in $D=4$}
In $D=4$, extremal   black-holes   preserving   one   supersymmetry  correspond  to
$N$-extended multiplets with
\begin{equation}
M_{ADM} = \vert Z_1 \vert >  \vert Z_2 \vert  \cdots > \vert Z_{[N/2]} \vert
         \end{equation}
         where $Z_\alpha$, $\alpha =1,\cdots, [N/2]$, are the proper values of
the central charge antisymmetric matrix written in normal form
 \cite{fesazu}.
The central charges $Z_{AB}= -Z_{BA}$, $A,B=1,\cdots,N$, and matter charges
 $Z_I$, $I= 1,\cdots , n$ are
those (moduli-dependent) symplectic invariant combinations of field strenghts
and their duals
(integrated over a large two-sphere)
 which appear
in the gravitino and gaugino supersymmetry variations respectively
\cite{cedafe}, \cite{noi1}, \cite{noi}.
Note that the total number of vector fields is $n_v=N(N-1)/2+n$ (with the
 exception of $N=6$
in which case there is an extra singlet graviphoton)\cite{cj}.
         \\
  It was shown in ref. \cite{feka2} that at the attractor point, where
  $M_{ADM}$ is extremized, supersymmetry requires that $Z_\alpha$, $\alpha >1$,
  vanish together with the matter charges $Z_I$, $I= 1, \cdots , n$
($n$ is the number of matter multiplets, which can exist only for $N=3,4$)
\par
This result can be used to show that for ``fixed scalars'',
corresponding to the attractor point, the scalar ``potential'' of the geodesic
 action \cite{bmgk}\cite{fegika}
\begin{equation}
V=-{1\over 2}P^t\cM(\cN)P
\label{sumrule1}
\end{equation}
is extremized in moduli space.
Here $P$ is the symplectic vector $P=(p^\Lambda, q_\Lambda) $ of quantized
electric and
magnetic charges and $\cM(\cN)$ is a symplectic $2n_v \times 2n_v$ matrix
 whose
$n_v\times n_v$ blocks are given in terms of the $n_v\times n_v$ vector
 kinetic matrix $\cN_{\Lambda\Sigma}$
($-Im \cN, Re \cN$ are the normalizations of the kinetic $F^2$ and the
topological $F^*F$
terms respectively) and
\begin{equation}
\cM(\cN) = \pmatrix{A& B \cr C & D \cr}
\end{equation}
with:
\begin{eqnarray}
A&=& Im \cN + Re \cN Im \cN^{-1} Re\cN \nonumber\\
B&=& - Re \cN Im \cN^{-1} \nonumber\\
C&=& -  Im \cN^{-1} Re\cN \nonumber\\
D&=&Im \cN^{-1}
\end{eqnarray}
The above assertion comes from the important identity, shown in ref.
 \cite{noi1}, \cite{noi} to be valid
in all $N\geq2$ theories:
\begin{equation}
-{1\over 2}P^t\cM(\cN)P = {1\over 2} Z_{AB} \bar Z^{AB} + Z_I \bar Z^I \label{sumrule2}
\end{equation}
Indeed, let us consider the differential relations satisfied by the charges
 \cite{noi}:
 \begin{eqnarray}
   \nabla Z_{AB} &=& {1\over 2} P_{ABCD} \bar Z^{CD} + P_{AB I} \bar Z^I \nonumber\\
 \nabla Z_{I} &=& {1\over 2} P_{AB I} \bar Z^{AB} + P_{IJ} \bar Z^J
 \label{differ}
 \end{eqnarray}
where the matrices $P_{ABCD}$, $P_{AB I}$, $P_{IJ}$ are the subblocks of the
 vielbein of $G/H$ embedded in $USp(n,n)$ \cite{noi}:
\begin{equation}
  \label{vielbein}
  \cP \equiv L^{-1} \nabla L = \pmatrix{P_{ABCD}& P_{AB J} \cr
P_{I CD} & P_{IJ} \cr }
\end{equation}
written in terms of the indices of $H=H_{Aut} \times H_{matter}$.
By computing the extremum of (\ref{sumrule1}) and using equations
(\ref{sumrule2}),(\ref{differ}) we obtain
\begin{equation}
\label{zz}
  P^{ABCD} Z_{AB} Z_{CD} =0 ;\quad Z_I=0
\end{equation}
$P_{ABCD}$ being the vielbein of the scalar manifold, completely antisymmetric
in its $SU(N)$ indices.
It is easy to see that in the normal frame these equations imply:
\begin{eqnarray}
  \label{norm}
  M_{ADM}\vert _{fix} & \equiv & \vert Z_{1} \vert \neq 0 \\
 \vert Z_{i} \vert & =& 0 \qquad (i=2, \cdots ,N/2 )
\end{eqnarray}

The main purpose of this section is to provide  particular expressions which
 give the
entropy formula as a moduli--independent quantity in the entire
moduli space and not just at the critical points.
Namely, we are looking for quantities $S\left(Z_{AB}(\phi), \bar Z^{AB}
 (\phi),Z_{I}(\phi), \bar Z^{I} (\phi)\right)$
such that ${\partial \over \partial \phi ^i} S =0$, $\phi ^i$ being the moduli
 coordinates.

These formulae generalize the quartic $E_{7(7)}$ invariant of $N=8$
supergravity \cite{kall} to all other cases.
\par
Let us first consider the theories $N=3,4$, where  matter can be
present \cite{maina}, \cite{bks}.
\par
The U-duality groups
 are, in these cases, $SU(3,n)$ and $SU(1,1)
\times SO(6,n)$ respectively (Here we denote by U-duality group the isometry
 group $G$
acting on the scalars, although only a restriction of it to integers is the
 proper U-duality group \cite{ht}).
The central and matter charges $Z_{AB}, Z_I$ transform in an obvious
way under the isotropy groups
\begin{eqnarray}
H&=& SU(3) \times SU(n) \times U(1) \qquad (N=3) \\
 H&=& SU(4) \times O(n) \times U(1) \qquad (N=4)
\end{eqnarray}
Under the action of the elements of $G/H$ the charges get mixed with
their complex conjugate.
\\
For $N=3$:
\begin{eqnarray}
P^{ABCD}&=& P_{IJ}=0 \, , \, \, P_{ AB I}  \equiv  \epsilon_{ABC}P^C_I \nonumber\\
\quad Z_{AB}
 &\equiv & \epsilon_{ABC}Z^C
 \label{viel3}
\end{eqnarray}
Then  the variations are:
\begin{eqnarray}
\delta Z^A  &=& \xi^A _I \bar Z^I  \\
\delta Z_{I}  &=&\xi^A _{ I} \bar Z_A
\label{deltaz3}
\end{eqnarray}
where $\xi^A_I$ are infinitesimal parameters of $K=G/H$.  Indeed,
once the covariant derivatives are known, the variations are obtained
by the  substitution $\nabla \to \delta$, $P \to \xi$.
\par
With a simple calculation, the U-invariant expression is:
\begin{equation}
S=   Z^A \bar Z_A - Z_I \bar Z^I
\label{invar3}
\end{equation}
In other words, $\nabla_i S = \partial_i S =0 $, where the covariant
derivative is
defined in ref. \cite{noi}.
\par
Note that at the attractor point ($Z_I =0$) it coincides with the
moduli-dependent potential (\ref{sumrule1})
computed at its extremum.
\\
For $N=4$
\begin{eqnarray}
P_{ABCD} &=& \epsilon_{ABCD}P ,\quad P_{IJ} = \eta_{IJ}\bar
P \nonumber\\
P_{AB I}&=& {1\over 2} \eta_{IJ} \epsilon_{ABCD}\bar P^{CD J} \label{viel4}
\end{eqnarray}
and the transformations of $K= {SU(1,1) \over U(1)} \times
{O(6,n) \over O(6) \times O(n)}$ are:
 \begin{eqnarray}
\delta Z_{AB}  &=& {1\over 2} \xi  \epsilon_{ABCD} \bar Z^{CD}  +
 \xi _{AB I} \bar Z^I \\
\delta Z_{I}  &=& \bar \xi \eta_{IJ} \bar Z^J + {1\over 2}\xi _{AB I} \bar Z^{AB}
\label{deltaz4}
\end{eqnarray}
with $\bar \xi^{AB I} =  {1\over 2} \eta^{IJ}   \epsilon^{ABCD}
\xi_{CD J}$.
\par
There are three $O(6,n)$ invariants given by $I_1$, $I_2$, $\bar I_2$ where:
\begin{eqnarray}
  I_1 &=& {1 \over 2}  Z_{AB} \bar Z_{AB} - Z_I \bar Z^I
\label{invar41} \\
I_2 &=& {1\over 4} \epsilon^{ABCD}  Z_{AB}   Z_{CD} - \bar Z_I \bar Z^I
\label{invar42}
\end{eqnarray}
and the unique   $SU(1,1)  \times
O(6,n) $  invariant $S$, $\nabla S =0$, is given by:
\begin{equation}
S= \sqrt{(I_1)^2 - \vert I_2 \vert ^2 }
\label{invar4}
\end{equation}
At the attractor point $Z_I =0$ and $\epsilon^{ABCD} Z_{AB} Z_{CD}
=0$, so that $S$ reduces to the square of the BPS mass.
\par
For $N=5,6,8$ the U-duality invariant expression $S$ is the square
root of a unique invariant under the corresponding U-duality groups
$SU(5,1)$, $O^*(12)$ and $E_{7(7)}$.
The strategy is to find a quartic expression $S^2$    in terms of
$Z_{AB}$ such that $\nabla S=0$, i.e. $S$ is moduli-independent.
\par
As before, this quantity is a particular combination of the $H$
quartic invariants.
\par
For $SU(5,1)$ there are only two  $U(5)$ quartic invariants.
In terms of the matrix $A_A^{\ B} = Z_{AC} \bar Z^{CB}$ they are:
$(Tr A)^2$, $Tr(A^2)$, where
\begin{eqnarray}
 Tr A & = & Z_{AB} \bar Z^{BA} \\
 Tr (A^2) & = & Z_{AB} \bar Z^{BC} Z_{CD} \bar Z^{DA}
\end{eqnarray}
As before, the relative coefficient is fixed by the transformation
properties of $Z_{AB}$ under ${SU(5,1) \over U(5) } $ elements of
infinitesimal parameter $\xi^C$:
\begin{eqnarray}
  \delta Z_{AB} = {1\over 2} \xi^C \epsilon_{CABPQ} \bar Z^{PQ}
\end{eqnarray}
It then follows that the required invariant is:
\begin{equation}
S= {1\over 2} \sqrt{  4 Tr(A^2) - (Tr A)^2 }
\label{invar5}
\end{equation}
For $N=8$ the $SU(8)$ invariants are:
\begin{eqnarray}
I_1 &=& (Tr A) ^2 \\
I_2 &=& Tr (A^2) \\
I_3 &=& Pf \, Z   \nonumber\\
 &=& {1\over 2^4 4!} \epsilon^{ABCDEFGH} Z_{AB} Z_{CD} Z_{EF} Z_{GH}
\end{eqnarray}
The ${E_{7(7)} \over SU(8)}$ transformations are:
\begin{equation}
\delta Z_{AB} ={1\over 2} \xi_{ABCD} \bar Z^{CD} \label{transf8}
\end{equation}
where $\xi_{ABCD}$ satisfies the reality constraint:
\begin{equation}
\xi_{ABCD} = {1 \over 24} \epsilon_{ABCDEFGH} \bar \xi^{EFGH}
\end{equation}
One finds the following $E_{7(7)}$ invariant \cite{kall}:
\begin{equation}
S= {1\over 2} \sqrt{4 Tr (A^2) - ( Tr A)^2 + 32 Re (Pf \,
Z) }
\label{quartinv}
\end{equation}
The $N=6$ case is the more complicated because under $U(6)$ the
left-handed spinor of $O^*(12)$ splits into:
\begin{equation}
32_L \to (15,1) + ( \bar {15}, -1) + (1, -3) + (1,3)
\end{equation}
The transformations of ${O^*(12) \over U(6)}$ are:
\begin{eqnarray}
\delta Z_{AB} &=& {1\over 4} \epsilon_{ABCDEF} \xi^{CD} \bar Z^{EF} +
\xi_{AB} \bar X \nonumber\\
\delta X &=& {1 \over 2} \xi _{AB} \bar Z^{AB} \label{transf6}
\end{eqnarray}
 where we denote by $X$ the $SU(6)$ singlet.
The quartic $U(6)$ invariants are:
\begin{eqnarray}
I_1&=& (Tr A)^2 \label{invar61}\\
I_2&=& Tr(A^2)\label{invar62} \\
I_3 &=& Re (Pf \, ZX) \nonumber\\
&=& {1\over 2^3 3!}
Re( \epsilon^{ABCDEF}Z_{AB}Z_{CD}Z_{EF}X)\label{invar63}\\
I_4 &=& (Tr A) X \bar X\label{invar64}\\
I_5&=& X^2 \bar X^2\label{invar65}
\end{eqnarray}
The unique $O^*(12)$ invariant is:
\begin{eqnarray}
S&=&{1\over 2} \sqrt{4 I_2 - I_1 + 32 I_3 +4I_4 + 4 I_5 }
\label{invar6}   \\
\nabla S &=& 0
\end{eqnarray}
Note that at the attractor point $Pf\,Z =0$, $X=0$ and $S$ reduces to
the square of the BPS mass.

We note that in the normal frame the transformations of the coset
which preserve the normal form of the $Z_{AB}$ matrix belong to
$O(1,1)^3$ both for $N=6$ and $N=8$ theories.
The relevant $O(1,1)^3$ transformations can be read out from eqs.
(\ref{transf8}), (\ref{transf6}) going to the normal frame.
The ensuing transformations correspond to commuting
matrices which are proper, non compact, Cartan elements of the coset
algebra of $N=8,N=6$ respectively\cite{uns}.

 \section{The attractor point condition in $D=5$}
In this section we  extend the previous analysis to the $D=5$
dimensional case,
for theories with $N>2$ supersymmetries.
Theories with $N=2$ at $D=5$ have been considered earlier \cite{feka1}
and fixed scalars recently analyzed in great detail \cite{kal}.
A technical important difference in this case is that although matter
charges vanish for fixed scalars,
preserving one  supersymmetry requires that the eigenvalues of the central
 charges which are not the BPS mass do not generally vanish at the horizon,
 but are all equal
and fixed in terms of the entropy.

The five dimensional case exhibits analogies and differencies with
respect to the four dimensional one.
Exactly like in the four dimensional theories,
the entropy is given, through the Bekenstein-Hawking relation,
by an invariant of the U-duality group over the entire moduli
space and its value is given in terms of the moduli-dependent
  scalar potential of the geodesic action\cite{bmgk} at the attractor point
  \cite{fks}, \cite{feka1},
  \cite{feka2}:
\begin{equation}
  \label{entropy}
  S= {A \over 4}={\pi ^2 \over 12} M^{3/2}_{extr}={\pi ^2 \over 12}
  \left[{\sqrt{3} \over 2}
  V(\phi _{fix}, g)\right]^{3/4}
\end{equation}
where we have used the relation $ V_{extr} = {4 \over 3} M^2_{extr}$
which is valid for any $N$ in $D=5$, as we will show in the following.
Note that, while in the four dimensional case the U-invariant is quartic
in the charges, in $D=5$
it turns out to be cubic.
Furthermore, in five dimensions the automorphism group under which the
central charges transform
is $USp(N)$ instead of $SU(N) \times U(1)$ as in the four dimensional
theories \cite{noi}, \cite{cj}.
As it is apparent from the dilatino susy transformation law, this implies that,
 at the minimum of the ADM mass, the central charges
 different from the maximal one do not vanish, contrary to what happens in
 $D=4$.
Indeed in $D=5$ we may perform again the extremization of the geodesic potential
as in $D=4$ but, due to the
traceless
condition of the antisymmetric symplectic representations of the vielbein
$P_{ABCD}$
and of the central charges $Z_{AB}$,the  analogous equations:
\begin{equation}
  \label{zz5}
  P^{ABCD} Z_{AB} Z_{CD} =0 ;\quad Z_I=0
\end{equation}
 do not imply anymore the
vanishing of the central charges different from the mass.
 Here $A,B,\cdots $  are $USp(N)$  indices
and  the antisymmetric matrix $Z_{AB}$ satisfies the reality condition:
\begin{equation}
{\bar Z}^{AB} = \IC^{AC}\IC^{BD}Z_{CD}
\end{equation}
$\IC^{AB}$ being the symplectic invariant antisymmetric matrix
satisfying $\IC = -\IC^T$, $ \IC^2 = - \bfone$.
Let us now consider more explicitly the various five dimensional theories.
In $N=4$ matter coupled supergravity \cite{noi}, the scalar manifold is given
by
 $G/H={O(5,n)\over O(5) \times O(n) } \times O(1,1)$.
The black-hole potential is given by:
\begin{eqnarray}
 V(\phi,q) &=& {1 \over 2} Z_{AB} \bar Z^{AB} + 2X^2 + Z_I Z^I \nonumber\\
 &=& q_\Lambda (\cN ^{-1})^{\Lambda\Sigma} q_\Sigma
\end{eqnarray}
where $X$ is the central charge associated to the singlet photon of
the $N=4$ theory,  $q_\Lambda \equiv \int_{S_3} \cN_{\Lambda\Sigma}
\,^\star F^\Sigma$ and $\cN _{\Lambda\Sigma}$
 are the electric charges
and vector kinetic matrix respectively.
The central charge $Z_{AB}$ can be decomposed in its $\IC$-traceless
 part $\buildrel \circ\over Z_{AB}$ and trace part $X$ according to:
\begin{equation}
  \label{centr4}
  Z_{AB} = { \buildrel \circ\over Z}_{AB} - \IC_{AB} X
\end{equation}
This decomposition corresponds to the combination of the five graviphotons
and the singlet photon
appearing in the gravitino transformation law \cite{noi}.
The matter charges $Z_I$ are instead in the vector representation of $O(n)$.
Note that in the dilatino transformation law a different
 combination of the five graviphotons and the singlet photon appears
corresponding to the integrated charge:
\begin{eqnarray}
  \label{zchi}
 Z^{(\chi)}_{AB}& \equiv & {1 \over 2}( Z_{AB} + 3 \, \IC_{AB}X ) \nonumber\\
 &=& {1 \over 2}({ \buildrel \circ\over Z}_{AB} + 2
 \, \IC_{AB}X )
\end{eqnarray}
The differential relations satisfied by the central and matter charges are:
\begin{eqnarray}
  \label{difrel4}
  \nabla Z_{AB} &=& P_{IAB} Z^I - 2 Z^{(\chi)}_{AB}d\sigma \nonumber\\
  & \to &
  \nabla {\buildrel \circ\over Z}_{AB} = P_{IAB} Z^I -
  {\buildrel \circ\over Z}_{AB}d\sigma \\
\nabla X & =& 2 X d\sigma \\
\nabla Z_I &=& {1\over 4}( Z_{AB}\bar P^{AB}_I \nonumber\\
&+& \bar Z^{AB}P_{AB I})
- Z_I d \sigma
\end{eqnarray}
To minimize the potential, it is convenient to go to the normal frame
where ${\buildrel \circ \over Z}_{AB}$ has proper values $e_1, e_2 = - e_1$.
In this frame, the potential and the differential relations become:
\begin{eqnarray}
V(\phi, q) &=& e_1^2 + e_2^2 + 4 X^2 \label{vnor4} \\
\nabla e_1 &=& - e_1 d \sigma  + P_{I} Z^I \\
 \nabla X & =& 2 X d\sigma
\end{eqnarray}
where $P_I \equiv P_{I 12}$ is the only independent component of the
traceless vielbein one-form $P_{I AB}$ in the normal frame.
We then get immediately that,
in the normal frame, the fixed scalar condition
${\partial V \over \partial \phi}=0$ requires:
\begin{eqnarray}
  Z_I&=& 0 \\
 {  e}_{1} &=&- { e}_{2}=  -2 X
  \end{eqnarray}
where ${ e}_{i}$ ($i=1,2$) are the proper  values of
${ \buildrel \circ\over Z}_{AB}$.
It follows that $M_{extr}= \vert Z_{12\ extr} \vert =  {3 \over 2} e_1$
so that
\begin{equation}
V_{extr}= 3 e_1^2 = {4 \over 3} M_{extr} ^2
\end{equation}

In the $N=6$ theory, the scalar manifold is $G/H = SU^\star (6)/Sp(6)$
\cite{noi}.
The central charge $Z_{AB}$ can again be  decomposed in a $\IC$-traceless
 part $\buildrel \circ\over Z_{AB}$ and a trace part $Z$ according to:
\begin{equation}
  \label{centr6}
  Z_{AB} = { \buildrel \circ\over Z}_{AB} + {1 \over 3} \, \IC_{AB} X
\end{equation}
The traceless and trace parts satisfy the differential relations:
\begin{eqnarray}
  \label{difrel6}
  \nabla { \buildrel \circ\over Z}_{AB} &=&
  { \buildrel \circ\over Z}_{C[A}P_{B]D} \, \IC^{CD}   \nonumber\\
  &+& {1\over 6}\, \IC_{AB}{ \buildrel \circ\over Z}_{LM}P^{LM}
  +{2 \over 3} XP_{AB}\\
\nabla X &=&{1\over 4} { \buildrel \circ\over Z}_{AB}P^{AB}
\end{eqnarray}
where $P_{AB}$ is the $\IC$ traceless vielbein of $G/H$.
The geodesic potential has the form:
\begin{eqnarray}
  \label{pot6}
  V(\phi,q)&=& {1\over 2} Z_{AB} Z^{AB} +  {4 \over 3}  X^2 \nonumber\\
  &=& q_{\Lambda\Sigma}
  (\cN ^{-1})^{\Lambda\Sigma , \Gamma\Delta}
  q_{\Gamma\Delta}
\end{eqnarray}
 where $ \Lambda,\Sigma , \cdots = 1, \cdots ,6$ are indices in
 the fundamental representation
 of $SU^\star (6)$, the couple of indices $\Lambda\Sigma$ in the
 elctric charges
 $q_{\Lambda\Sigma}$ are antisymmetric and $ (\cN )_{\Lambda\Sigma
 , \Gamma\Delta} $ is
 the kinetic matrix  of the vector field-strengths
 $F^{\Lambda\Sigma}$
To perform the minimization of the potential we proceed as in the
$N=4$ case going to the normal frame where ${ \buildrel \circ\over Z}_{AB}$
has proper values  $e_1, e_2, e_3=-e_1-e_2$.
The potential becomes:
  \begin{equation}
  \label{potnor6}
  V(\phi,q)=  e_1^2 + e_2^2 + (e_1 + e_2 )^2 + {4 \over 3}X^2
\end{equation}
   Moreover, the differential relations (\ref{difrel6}) take the
   form:
 \begin{eqnarray}
\nabla e_1 &=& {1\over 3} (- e_1+ e_2 + 2 X)P_1 \nonumber\\
&+& {1\over 3}(e_1 + 2
e_2) P_2 \label{difnor61}\\
 \nabla e_2 &=&   {1\over 3}(2e_1 +
e_2) P_1  \nonumber\\
&+& {1 \over 3} ( e_1 - e_2 + 2 X)P_2 \label{difnor62}\\
 \nabla X &=& {1 \over 2} ( 2e_1 + e_2)P_1 + {1 \over 2} (e_1 + 2
e_2) P_2
\label{difnor63}
\end{eqnarray}
 where $P_1,P_2,P_3=-P_1 -
P_2$ are the proper values of the vielbein one-form
$P_{AB}$ in the normal frame.
Imposing the attractor-point constraint ${\partial V \over \partial \phi} =0$
on the potential we get the following
relations among the charges at the extremum:
\begin{eqnarray}
 { e}_{2} &=&   { e}_{3} =  - {1\over 2} {  e}_{1}=  - {4 \over 3}X
 \nonumber\\
   V_{extr} &=& {27 \over 16} e_1^2 = {4 \over 3} M_{extr}^2 .
 \label{min6}
  \end{eqnarray}
Note that, using eqs. ( \ref{difnor61})-(\ref{difnor63}), (\ref{min6}),
the mass $e_1 + {1 \over 3} X $
satisfies ${ \partial \over \partial \phi^i} (e_1 + {1 \over 3} X  )
=0$ at the extremum,
with value $M_{extr} = {9 \over 8} e_1$.
\vskip 0.5cm
In the $N=8$ supergravity the scalar manifold is $G/H = E_{6(-6)}/Sp(8)$
and the central charges sit in the  twice antisymmetric, $\IC$-traceless,
representation of $USp(8)$ \cite{noi}.
The scalar ``potential'' of the geodesic action is given by:
\begin{eqnarray}
  \label{pot8}
  V(\phi , g) &=& {1 \over 2} Z_{AB} \bar Z^{AB} \nonumber\\
  &=& q_{\Lambda\Sigma}
  (\cN^{-1}) ^{\Lambda\Sigma, \Gamma\Delta}(\phi ) q_{\Gamma\Delta}
\end{eqnarray}
where $q_{\Lambda\Sigma} \equiv \int\cN_{\Lambda\Sigma ,
\Gamma \Delta}F^{\Gamma\Delta}$
are the electric charges
 and $\cN _{\Lambda\Sigma, \Gamma\Delta}$ the vector kinetic matrix.
The extremum of $V$ can be found by using the differential relation
\cite{noi}:
\begin{equation}
  \label{difrel}
  \nabla Z_{AB} ={1 \over 2} P_{ABCD} \bar Z^{CD}
\end{equation}
where $P_{ABCD}$ is the four-fold antisummetric vielbein one-form of
$G/H$.
One obtains:
\begin{equation}
 P^{ABCD} Z_{AB} Z_{CD} =0
\end{equation}

To find the values of the charges at the extremum we use the
traceless conditions
$\IC^{AB} P_{ABCD} =0$, $\IC^{AB} Z_{AB} =0$.
In the normal frame, the proper values of $Z_{AB}$ are
$e_1, e_2, e_3, e_4=-e_1 -e_2- e_3$  and we take, as
independent components of the vielbein,
$P_1 = P_{1234} = P_{5678}$, $P_2 = P_{1256}  = P_{3478}$ ($P_{3456}=
P_{1278} = - P_1 - P_2$).
In this way, the covariant derivatives of the charges become:
\begin{eqnarray}
\label{difnor8}
\nabla e_1 &=& (e_1 + 2 e_2 + e_3) P_1 \nonumber\\
&+& (e_1 + e_2 + 2 e_3) P_2 \\
\nabla e_2 &=& (e_1 - e_3) P_1 + (-e_1 - e_2 - 2 e_3) P_2 \\
\nabla e_3 &=& (-e_1 - 2 e_2 - e_3) P_1 + (e_1 - e_2) P_2
\end{eqnarray}
Using these relations, the extremum condition of $V$ implies:
\begin{eqnarray}
  \label{extrz}
  e_2 &=& e_3 = e_4 = -{1\over 3} e_1 \nonumber\\
  V_{extr} &=& {4 \over 3} e_1^2 = {4 \over 3} M_{extr}^2
\end{eqnarray}

\section{Topological invariants}
In this section we determine the U-invariant expression in terms of
which the entropy can be evaluated over the entire moduli space.
Our procedure is the same as in ref. \cite{uns}, namely we compute
cubic H-invariants
and determine the appropriate linear combination of them which turns
out to be U-invariant.
The invariant expression of the entropy for $N=4$ and $N=8$ at $D=5$
in terms of the quantized charges was given in \cite{feka2}.
Let us begin with the $N=4$ theory, controlled by the coset
${O(5,n)\over O(5)\times O(n)} \times O(1,1)$.
In this case there are three possible cubic H-invariants, namely:
 \begin{eqnarray}
I_1 &=& {1\over 2} {\buildrel \circ \over Z}_{AB}  {\buildrel \circ
\over {\bar Z}}^{AB}X   \\
 I_2 &=& Z_I Z_I X \\
I_3 &=& X^3
\end{eqnarray}
In order to determine the $U \equiv G= O(5,n) \times O(1,1)$-invariant,
we set $I=I_1 + \alpha I_2 + \beta I_3$ and
  using the differential  relations  (\ref{difrel4}) one easily
finds that $\partial I =0$ implies $\alpha =1$, $\beta =0$.
Therefore:
\begin{equation}
I= I_1 - I_2 = \left({1\over 2} \buildrel\circ\over Z_{AB}
{\buildrel\circ\over {\bar Z}}^{AB} - Z_I Z_I \right)X
\end{equation}
 is the cubic ($O(5,n) \times O(1,1)$)-invariant independent
of the moduli and the entropy $S={\pi \over 12} M^{3\over 2}$ is
then given as:
\begin{equation}
S \sim I^{1/2}= \sqrt{({1\over 2} Z_{AB}\bar Z^{AB} -Z_I Z_I)Z }
\label{cubinv4}
\end{equation}
In the $N=6$ theory, where the coset manifold is $SU^*(6) /Sp(6)$,
the possible cubic $Sp(6)$-invariants are:
\begin{eqnarray}
I_1&=& Tr(Z\IC)^3 \\
 I_2&=& Tr(Z\IC)^2X \\
 I_3&=& X^3
\end{eqnarray}
Setting as before:
\begin{equation}
I= I_1 + \alpha I_2 + \beta I_3
\end{equation}
the covariant derivative $\partial I$ is computed using the differential
relations (\ref{difrel6}) and the parametrer
$\alpha$ and $\beta$ are then determined by imposing $\partial I =0$.
Actually, the best way to perform the computation is to go to the
normal frame.
Using the differential relations (\ref{difnor61})-(\ref{difnor63})
and the expression for the invariants in the normal frame:
\begin{eqnarray}
Tr(Z\IC)^3 &=& - 6 (e_1^2 e_2 + e_1 e_2 ^2)\\
Tr(Z\IC)^2 &=& 4 (e_1^2+ e_2^2 + e_1 e_2 )
\end{eqnarray}
  the vanishing of $\partial I$ fixes the coefficients
$\alpha$ and $\beta$.
The final result is:
\begin{equation}
I=-{1\over 6}Tr( Z\IC)^3 - {1\over 6} Tr( Z\IC)^2 X + {8 \over 27} X^3
\label{cubinv6}
\end{equation}
and the entropy is:
\begin{equation}
S \sim \sqrt{ -{1\over 6}Tr( Z\IC)^3 -  {1\over 6}Tr( Z\IC)^2 X +
{8 \over 27} X^3}
\label{cubinv8}
\end{equation}

Finally, for the $N=8$ theory, described by the coset
$E_{6(6)}/Sp(8)$,  it is well known that there is  a unique $E_{6}$
cubic invariant, namely:
\begin{equation}
I_3(27)=Tr(Z\IC )^3= Z_{A}^{\ B} Z_{B}^{\ C} Z_{C}^{\ A}
\label{i327}
\end{equation}
where the $Sp(8)$ indices are raised and lowered by the antisymmetric
matrix $\IC_{AB}$.
Curiously, the $E_6$-invariant corresponds to a single cubic
$USp(8)$-invariant.
Again, the invariance of $I$ can be best computed in the normal
frame where the   invariant (\ref{cubinv8})  takes the form:
\begin{eqnarray}
I_3(27)&=& Tr(Z\IC )^3 = -2(e_1^3 + e_2 ^3 + e_3^3 + e_4 ^3) \nonumber\\
 &=& 6(e_2^2 e_3+ e_2
 e_3 ^2 + e_1^2e_2 + e_1e_2^2\nonumber\\
 &+& e_1^2e_3 + e_1e_3^2+ 2 e_1e_2 e_3)
\label{i3}
\end{eqnarray}
One finds indeed $\partial I=0$ and therefore
\begin{equation}
S \sim I^{1/2} = \sqrt{Tr(Z\IC)^3}
\end{equation}
As in $D=4$ the transformations of the coset
which preserve the normal form of the $Z_{AB}$ matrix belong to the
non compact Cartan subalgebra of $SU^*(6)$ and $E_{6(6)}$ for $N=6$ and $N=8$ respectively, which in
both cases turns out to be $SO(1,1)^2$.

The relevant $O(1,1)^2$ transformations can be read out from eqs.
(\ref{difnor61})-(\ref{difnor63}), (\ref{difnor8}) written in the
normal frame\cite{5d}.


\section{BPS Conditions for Enhanced Supersymmetry}
In this section we will describe U-duality invariant constraints
on the multiplets of quantized charges in the case of BPS black holes
 whose background preserves more than one supersymmetry \cite{fema}.

We will still restrict our analysis to four and five dimensional cases
for which three possible cases exist {\it i.e.} solutions preserving
 $1/8$, $1/4$ and $1/2$ of the original  supersymmetry (32 charges).

The invariants may only be non zero on solutions preserving $1/8$ supersymmetry.
In dimensions $6\le D \le 9$ black holes may only preserve $1/4$ or $1/2$
supersymmetry, and no associated invariants exist in these cases.

The description which follows also make contact with the D-brane
 microscopic calculation, as it will appear
obvious from the formulae given below.
We will first consider the five dimensional case.

In this case, BPS states preserving $1/4$  of supersymmetry correspond
 to the invariant constraint $I_3 (27) =0$ where $I_3$ was defined in eq. \ref{i327}.
This corresponds to the $E_6$ invariant statement that the {\bf 27} is a null vector
with respect to the cubic norm.
As we will show in a moment, when this condition is fulfilled
 it may be shown that two of the central charge eigenvalues
are equal in modulus.
The generic configuration has 26 independent charges.

Black holes corresponding to $1/2$ BPS states correspond to null vectors
which are critical, namely
\begin{equation}
\partial I(27) =0
\end{equation}

In this case the three central charge eigenvalues are equal in modulus
and a generic charge vector has 17 independent components.

To prove the above statements, it is useful to compute the cubic invariant
in the normal frame, given  by:
\begin{eqnarray}
I_3(27) &=& Tr(Z\IC )^3 \nonumber\\
&=& 6 (e_1+e_2)(e_1+e_3)(e_2+e_3)\nonumber\\
& = & 6 s_1s_2s_3
\end{eqnarray}
where:
\begin{eqnarray}
e_1 &=& {1 \over 2} (s_1 + s_2 - s_3 ) \nonumber\\
e_2 &=& {1 \over 2} (s_1 - s_2 + s_3 ) \nonumber\\
e_3 &=& {1 \over 2} (- s_1 + s_2 + s_3 )
\end{eqnarray}
are the eigenvalues of the traceless antisymmetric $8\times 8$
matrix.
We then see that if $s_1=0$ then $\vert e_1\vert = \vert e_2\vert $,
and if $s_1= s_2 =0$ then $\vert e_1\vert = \vert e_2\vert = \vert e_3\vert $.
To count the independent charges we must add to the eigenvalues
the angles given by $USp(8)$ rotations.
The subgroup of $USp(8)$ leaving two eigenvalues invariant is $USp(2)^4$,
which is twelve dimensional.
The subgroup of $USp(8)$ leaving invariant one eigenvalue is $USp(4) \times USp(4)$,
which is twenty dimensional.
The angles are therefore $36-12=24$ in the first case, and $36-20=16$ in the second case.
This gives rise to configurations with 26 and 17 charges respectively, as promised.

Taking the case of Type II on $T^5$ we can choose $s_1$ to correspond
to a solitonic five-brane charge, $s_2$ to a fundamental string
winding charge along some direction and $s_3$ to Kaluza-Klein
momentum along the same direction.

The basis chosen in the  above example is S-dual to the D-brane basis
usually chosen for describing black holes in Type IIB on $T_5$.
All other bases are related by U-duality to this particular choice.
We also observe that the above analysis relates the cubic invariant to the picture  of
intersecting branes since a three-charge $1/8$ BPS configuration with non vanishing entropy
  can be thought as obtained by intersecting three single charge $1/2$ BPS configurations
\cite{bdl}, \cite{bll}, \cite{lptx}

By using the S--T-duality decomposition we see that the cubic
invariant reduces to $I_3(27)=10_{-2} 10_{-2} 1_4 + 16_1 16_1
10_{-2}$.
The 16 correspond to D-brane charges, the 10 correspond to the 5 KK
directions and winding of wrapped fundamental strings, the 1
correspond to the N-S five-brane charge.

We see that to have a non-vanishing area we need a configuration with
three non-vanishing N-S charges or two D-brane charges and one N-S charge.

Unlike the 4-$D$ case, it is impossible to have a non-vanishing entropy
for a configuration only carrying D-brane charges.

We now turn to the four dimensional case.

In this case the situation is more subtle because the condition for the 56
to be a null vector (with respect to the quartic norm defined
through eq. \ref{quartinv}) is not sufficient to enhance the supersymmetry.
This can be easily seen by going to the normal frame where it can be shown
that for a null vector there are not, in general,
coinciding eigenvalues.
The condition for $1/4$ supersymmetry is that the gradient of the quartic invariant vanish.

The invariant condition for $1/2$ supersymmetry is that the second
 derivative projected into the adjoint representation
of $E_7$ vanish.
This means that, in the symmetric quadratic polynomials of second derivatives,
only terms in the 1463 of $E_7$ are non-zero.
Indeed, it can be shown, going to the normal frame for the 56 written as a skew
$8 \times 8$ matrix, that the above conditions imply two and four eigenvalues
 being equal respectively.

The independent charges of $1/4$ and $1/2$ preserving supersymmetry
are 45 and 28 respectively.

To prove the latter assertion,
it is sufficient to see that the two charges normal-form matrix is left
 invariant by $USp(4) \times USp(4)$,
while the one charge matrix is left invariant by $USp(8)$ so the $SU(8)$ angles are
$63-20=43$ and $63-36=27$ respectively.

The generic $1/8$ supersymmetry preserving  configuration of the 56 of $E_7$
with non vanishing
entropy has five independent parameters in the normal frame and $51=63-12$
$SU(8)$ angles.
This is because the compact little group of the normal frame is $SU(2)^4$.
The five parameters describe the four eigenvalues and an overall phase
of an $8\times 8$ skew diagonal matrix.

If we allow the phase to vanish, the 56 quartic norm just simplifies
as in the five dimensional case:
\begin{eqnarray}
&& I_4(56) = s_1s_2s_3s_4 \nonumber\\
&&= (e_1+e_2+e_3+e_4)(e_1+e_2-e_3-e_4)\nonumber\\
&&\times (e_1-e_2-e_3+e_4)(e_1-e_2+e_3-e_4)
\label{i4inv}
\end{eqnarray}
where $e_i$ ($i=1,\cdots, 4$) are the four eigenvalues.

$1/4$ BPS states correspond to $s_3=s_4=0$ while $1/2$ BPS states correspond to
$s_2=s_3=s_4=0$.

An example of this would be a set of four D-branes oriented along 456, 678, 894, 579
(where the order of the three numbers indicates the orientation of the brane).
Note that in choosing the basis the sign of the D-3-brane charges is important; here they are chosen
such that taken together with positive coefficients they form a BPS object.
The first two possibilities  ($I_4 \neq 0$ and $I_4 =0$,
 ${\partial I_4 \over \partial q^i \neq 0}$) preserve $1/8$ of the supersymmetries, the
third (${\partial I_4 \over \partial q^i} = 0$, ${\partial^2 I_4 \over \partial q^i\partial q^j}\vert_{Adj~E_7} \neq 0$)
$1/4$ and the last (${\partial^2 I_4 \over \partial q^i\partial q^j}\vert_{Adj~E_7} = 0$) $1/2$.

It is interesting that there are two types of $1/8$ BPS solutions.
In the supergravity description, the difference between them is that the first case has non-zero
horizon area.
If $I_4 <0$ the solution is not BPS.
This case corresponds, for example, to changing the sign of one of the three-brane
charges discussed above.
By U-duality transformations we can relate this to configurations of branes at angles such as in \cite{balale}

Going from four to five dimensions it is natural to decompose the $E_7 \to E_6 \times O(1,1)$
where $E_6$ is the duality group in five dimensions and $O(1,1)$ is
 the extra T-duality that appears when we compactify from five to four
dimensions.
According to this decomposition, the representation breaks as: ${\bf 56} \to {\bf 27}_1 + {\bf 1}_{-3}
+ {\bf 27}'_{-1} + {\bf 1}_3$ and the quartic invariant becomes:
\begin{eqnarray}
  \label{decomp}
  {\bf 56}^4 &=& ({\bf 27}_1)^3{\bf 1}_{-3}+({\bf 27^\prime}_{-1})^3{\bf 1}_{3} +
  {\bf 1}_{3} {\bf 1}_{3}{\bf 1}_{-3}{\bf 1}_{-3}\nonumber\\
&+& {\bf 27}_1{\bf 27}_1{\bf 27}'_{-1}{\bf 27}'_{-1} +{\bf 27}_1{\bf 27}'_{-1} {\bf 1}_{3}{\bf 1}_{-3}
\end{eqnarray}
The ${\bf 27}$ comes from point-like charges in five dimensions and the ${\bf 27}'$
comes from string-like charges.

Decomposing the U-duality group into T- and S-duality groups, $E_7 \to SL(2,\IR)\times O(6,6)$,
we find ${\bf 56} \to ({\bf 2,12}) + ({\bf 1,32})$ where the first term
corresponds to N-S charges and the second term to D-brane charges.
Under this decomposition the quartic invariant (\ref{i4inv}) becomes
${\bf 56}^4 \to {\bf 32}^4 + ({\bf 12 . 12'})^2 + {\bf 32}^2.{\bf 12 . 12'}$.
This means that we can have configurations with a non-zero area that carry
only D-brane charges, or only N-S charges or both D-brane and N-S charges.

It is remarkable that $E_{7(7)}$-duality gives additional restrictions on the BPS states
other than the ones merely implied by the supersymmetry algebra.
The analysis of double extremal black holes implies that $I_4$ be semi-definite positive
for BPS states.
From this fact it follows that configurations preserving $1/4$ of supersymmetry must have eigenvalues
equal in pairs, while configuratons with three coinciding eigenvalues are not BPS.

To see this,  it is sufficient to write the quartic invariant in the normal frame
basis.
A generic skew diagonal $8\times 8$ matrix depends on four complex eigenvalues $z_i$.
These eight real parameters can be understood using the decomposition \cite{trig}:
\begin{eqnarray}
  {\bf 56} &\to & ( {\bf 8_v,2,1,1}) + ({\bf 8_s,1,2,1})+ ({\bf 8_c,1,1,2})\nonumber\\
 &+& ({\bf 1,2,2,2})
\end{eqnarray}
under
\begin{equation}
  E_{7(7)} \to O(4,4) \times SL(2,\IR)^3
\end{equation}
Here $O(4,4)$ is the little group of the normal form and the
 $({\bf 2,2,2})$ are the four complex skew-diagonal elements.
We can further use $U(1)^3 \subset SL(2,\IR)^3$ to further remove three relative phases
so we get the five parameters $z_i = \rho_i e^{{\rm i} \phi/4}$ ($i=1,\cdots, 4$).

The quartic invariant, which is also the unique $SL(2,\IR)^3$ invariant built with the $({\bf 2,2,2})$,
becomes \cite{fema}:
\begin{eqnarray}
 &&I_4= \sum_i \vert z_i \vert^4 -2 \sum_{i<j} \vert z_i \vert^2 \vert z_j \vert ^2
 \nonumber\\
&&+ 4 (z_1z_2z_3z_4 + {\bar z_1}{\bar z_2}{\bar z_3}{\bar z_4} )\nonumber\\
&&= (\rho_1 + \rho_2 + \rho_3 + \rho_4) (\rho_1 + \rho_2 - \rho_3 - \rho_4)\nonumber\\
\times
&&(\rho_1 - \rho_2 + \rho_3 + \rho_4)(\rho_1 - \rho_2 - \rho_3 + \rho_4)\nonumber\\
&&+8\rho_1\rho_2\rho_3\rho_4(cos \phi -1)
\end{eqnarray}

The last term is semi-definite negative.
The first term, for $\rho_1=\rho_2=\rho$ becomes:
\begin{equation}
  -[4\rho^2 -(\rho_3+\rho_4)^2](\rho_3-\rho_4)^2
\end{equation}
which is negative unless $\rho_3=\rho_4$.
So $1/4$ BPS states must have
\begin{equation}
\rho_1=\rho_2 > \rho_3=\rho_4~,\quad cos \phi =1
\end{equation}
For $\rho_1=\rho_2 =\rho_3=\rho$, the first term in $I_4$ becomes:
\begin{equation}
  -(3\rho + \rho_4)(\rho - \rho_4)^3
\end{equation}
so we must also have $\rho_4=\rho~,\quad cos \phi=1$
which is the $1/2$ BPS condition.

An interesting case, where $I_4$ is negative, corresponds to a configuration carrying
electric and magnetic charges under the same gauge group, for example a 0-brane
plus 6-brane configuration which is dual to a K--K-monopole plus K--K-momentum
 \cite{ko}, \cite{shein}.
This case corresponds to $z_i=\rho e^{{\rm i} \phi/4}$ and the phase is $tan \phi/4 =e/g$
where $e$ is the electric charge and $g$ is the magnetic charge.
Using (\ref{i4inv}) we find that $I_4<0$ unless the solution is purely electric or
purely magnetic.
In \cite{pol} it was suggested that  $0+6$ does not form a supersymmetric state.
Actually, it was shown in \cite{tay} that a $0+6$ configuration can be
 T-dualized into a non-BPS configuration of four intersecting D-3-branes.
Of course, $I_4$ is negative for both configurations.
Notice that even though these two charges are Dirac dual (and U-dual)
they are not S-dual in the sense of filling out an $SL(2,\ZZ)$
multiplet.
In fact, the K--K-monopole forms an $SL(2,\ZZ)$ multiplet with a fundamental string winding
charge under S-duality \cite{sen}

\section{Duality Orbits fo BPS States Preserving Different Numbers of Supersymmetries}
In this section we give an invariant classification of BPS black holes preserving
 different numbers of supersymmetries in terms
of orbits of the {\bf 27} and the {\bf 56} fundamental representations of the duality
groups $E_{6(6)}$ and $E_{7(7)}$ resperctively \cite{fg}, \cite{stelle}.

In five dimensions the generic orbits preserving $1/8$ supersymmetry correspond to the
26 dimensional orbits $E_{6(6)}/F_{4(4)}$
so we may think the generic 27 vector of $E_6$ parametrized by a point in this orbit
and its cubic norm (which actually equals the square of the black-hole entropy).

The light-like orbit, preserving $1/4$ supersymmetry, is the 26
dimensional coset $E_{6(6)}/O(5,4) \odot T_{16}$
where $ \odot $ denotes the semidirect product.

The critical orbit, preserving maximal $1/2$ supersymmetry (this correspond to
${\partial I_4 \over \partial q^i }\neq 0$) correspond to the 17 dimensional space
\begin{equation}
  {E_{6(6)} \over O(5,5) \odot T_{16}}
\end{equation}
In the four dimensional case, we have two inequivalent 55 dimensional orbits
corresponding to the cosets $E_{7(7)}/E_{6(2)}$ and $E_{7(7)}/E_{6(6)}$ depending
on whether $I_4(56)>0$ or $I_4(56)<0$.
The first case corresponds to $1/8$ BPS states whith non-vanishing entropy,
while the latter corresponds to non BPS states.

There is an additional 55 dimensional light-like orbit ($I_4 =0$) preserving
$1/8$ supersymmetry given by ${E_{7(7)} \over F_{4(4)}\odot T_{26}}$.

The critical light-like orbit, preserving $1/4$ supersymmetry, is the
45 dimesnsional coset  $E_{7(7)}/O(6,5) \odot (T_{32}\oplus T_1)$

The critical orbit corresponding to maximal $1/2$ supersymmetry is described
by the 28 dimensional quotient space
\begin{equation}
{E_{7(7)} \over E_{6(6)}\odot T_{27}}
\end{equation}
We actually see that the counting of parameters in terms of invariant
orbits reproduces the counting previously made in terms
of normal frame parameters and angles.
The above analysis makes a close parallel between BPS states preserving different numbers of
supersymmetries with time-like, space-like and light-like vectors in Minkowski space.

\section*{Acknowledgements}
The results of sections 5 and 6 have been obtained  by one of the
authors (S. F.) in collaborations with J.M. Maldacena and M. Gunaydin.


\end{document}